\documentclass[prl,twocolumn,superscriptaddress,preprintnumbers,amsmath,amssymb,floatfix,nofootinbib]{revtex4}

\usepackage{graphicx}
\usepackage{amsmath}    
\usepackage[normalem]{ulem}
\usepackage{bm}

\usepackage[utf8x]{inputenc}
\usepackage{multirow}
\usepackage{subfigure}
\usepackage{booktabs}
\usepackage{epsfig,graphics}
\usepackage{amsfonts,amssymb,amsmath, mathrsfs}            
\usepackage{amsthm}

\newcommand{\nc}{\newcommand}

\nc{\be}{\begin{equation}}

\nc{\ee}{\end{equation}}

\nc{\bea}{\begin{eqnarray}}

\nc{\eea}{\end{eqnarray}}

\nc{\xx}{\nonumber\\}

\nc{\ct}{\cite}

\nc{\la}{\label}

\nc{\eq}[1]{(\ref{#1})}

\def\ajou#1&#2(#3){\ \sl#1\bf#2\rm(19#3)}

\def\[{\left [}
\def\]{\right ]}

\theoremstyle{plain}

\theoremstyle{definition}

\begin{document}
\title{First Integrals of Generalized Darboux-Halphen Systems and Membrane Paradigm}

\author{Sumanto Chanda}
\affiliation{S.N. Bose National Centre for Basic Sciences \\ JD Block, Sector-3, Salt Lake, Calcutta-700098, INDIA}

\author{Partha Guha}
\affiliation{S.N. Bose National Centre for Basic Sciences \\ JD Block, Sector-3, Salt Lake, Calcutta-700098, INDIA}

\author{Raju Roychowdhury}
\affiliation{Instituto de Fisica, Universidade de Sao Paulo, C. Postal 66318, 05314-970 Sao Paulo, SP, Brazil}

\date{\today}

\begin{abstract}

The Darboux-Halphen system of equations have common or individual additive terms 
depending on the matrices defining Yang-Mills gauge potential fields. 
Tod  \cite{Tod}, described a conserved quantity for the classical 
systems with no additive terms. We show that the conserved quantity apply even for
the generalized cases with common additive terms. A theory has been presented, with an example, 
of how to formulate conserved quantities for equation with individual additive terms.
We also briefly shed some light on the issues of surface motions of fluids in connection to Nahm`s equation
 and the  self-duality and integrability of membrane dynamics.

\end{abstract}


\maketitle

\section{Introduction}

The Darboux-Halphen differential equations often referred to as the classical Darbou-Halphen system

\vspace{-0.5cm}
\begin{equation}
\label{DH} \dot{\omega}_i = \omega_j \omega_k - \omega_i \big( \omega_j + \omega_k \big) \hspace{1cm} i \neq j \neq k = 1, 2, 3
\end{equation}

was originally formulated by Darboux \cite{Da} and subsequently solved by Halphen \cite{Ha}. The general solution to equation (\ref{DH}) may be expressed in terms of the elliptic modular function. In fact Halphen related the DH equation with the Legendre type hypergeometric equation.

The system (\ref{DH}) has found applications in mathematical physics in relation to magnetic monopole dynamics \cite{AH}, self dual Einstein equations \cite{GP, Hi}  and topological field theory \cite{Du}. Ablowitz et al \cite{AC} studied the  reduction  of  the  self-dual Yang-Mills equation (SDYM)  with  an  infinite-dimensional  Lie  algebra  to  a  generalized  Darboux-Halphen  system  whose  general  solution  is densely  branched  about  movable  singularities  and  can  contain  movable  natural  barriers. However, the fully generalized Darboux-Halphen system with individual additive terms has not been formulated, aside from an example in \cite{AC}.

Recently in \cite{cgr}, the Darboux-Halphen system was reviewed from the perspective of the self-dual Bianchi-IX metric and the self-dual Yang-Mills field equation, describing a gravitational instanton in the former case, and a Yang-Mills instanton in the latter. All systems related to the Darboux-Halphen system such as Ramanujan and Ramamani system were covered, as well as aspects of integrability of the Darboux-Halphen system. The analysis was restricted to the classical Darboux-Halphen system leaving the generalized version undiscussed. 

Tod \cite{Tod}, found explicit expressions for a family of self-dual Bianchi-IX metrics that were obtained from the Painleve-VI equation. In those cases where the anti-self-dual Weyl tensor and scalar curvature vanish, the system was described by the classical Darboux-Halphen system. Many properties related to the system including the conserved quantity, the Darboux-Halphen equations and the Bianchi-IX metric under reparametrization were discussed therein. 

In this article, we will show that the conserved quantity also applies to the generalized Darboux-Halphen system with common additive terms. Then we shall proceed to formulate generalized equations with individual additive terms, not formulated anywhere aside from the example found by Ablowitz and Chakravarty in \cite{AC}. After that, we shall deduce the conserved quantity for this example.

First, we shall derive the conserved quantity or first integral $Q$ related to the generalized Darboux-Halphen equations, which applies to the classical case simultaneously. It is followed by a rewriting of Tod's description of a change of variables redefining the conserved quantity. The following section covers how we can obtain further generalized Darboux-Halphen equations, where we will have individual additive terms to each equation that describes a general formulation of conserved quantity in this case as well. the penultimate section explores integrable membrane dynamics with an explicit construction of Lax pair 
for the Nahm equation obtained ala reduction from the self-dual Yang-Mills equation. Finally we conclude and pave the path for future directions.

\section{Generalized DH system with common additive terms}

In this section, we show that Tod's results for the conserved quantity related to the classical system can be extended to the generalized Darboux-Halphen system with common additive terms \cite{fintgradflo}, described by the equations:

\vspace{-0.5cm}
\begin{equation}
\label{gendh} \dot{\omega}_i = \omega_j \omega_k - \omega_i \big( \omega_j + \omega_k \big) + \lambda^2 \hspace{0.5cm} i \neq j \neq k = 1, 2, 3
\end{equation}

with the common additive term $\lambda^2$ elaborated as

\vspace{-0.25cm}
\begin{equation}
\label{diff}
\begin{split}
\lambda^2 &= \alpha_1^2 x_2 x_3 + \alpha_2^2 x_3 x_1 + \alpha_3^2 x_1 x_2 \\  
\text{where } &\hspace{0.5cm} x_i = \omega_j - \omega_k \hspace{0.75cm} i \neq j \neq k = 1, 2, 3
\end{split}
\end{equation}

The generalized Darboux-Halphen system arises from the self-dual Yang-Mills equation \cite{ACH}, as a result of solving Nahm's equations. Using the three equations described by (\ref{gendh}), we will now deduce the conserved quantity or first integral for systems with shared additive terms.

\subsection{The conserved quantity of the Generalized Darboux-Halphen systems}

To obtain a conserved quantity, we proceed with (\ref{gendh}) as we would with the classical system (\ref{DH}). For this, an important equation for the variable $x_i$ defined in (\ref{diff}) is:

\vspace{-0.5cm}
\begin{equation} \label{xprop1} 
\dot{x}_i = - 2 \omega_i x_i \hspace{1cm} x_1 + x_2 + x_3 = 0
\end{equation}

Thus, starting with (\ref{gendh}) and using (\ref{xprop1}), we will have

\vspace{-0.5cm}
\begin{equation}
\label{termi} \bigg( \frac{\omega_i^2}{x_j x_k} \bigg)' = 2 \frac{\omega_1 \omega_2 \omega_3}{x_1 x_2 x_3} x_i - \frac{\lambda^2}{x_1 x_2 x_3} \dot{x}_i
\end{equation}

On adding up these terms (\ref{termi}) over $i$ keeping $i < j < k$, and applying the sum rule of (\ref{xprop1}), we find that the result is a vanishing derivative:

\vspace{-0.5cm}
\begin{equation}
\sum_i^{ i < j < k} \bigg( \frac{\omega_i^2}{x_j x_k} \bigg)' = \frac{\omega_1 \omega_2 \omega_3}{x_1 x_2 x_3} \sum_i x_i - \frac{\lambda^2}{x_1 x_2 x_3} \sum_i \dot{x}_i = 0 \nonumber
\end{equation}

Thus, it is clear that even for the generalized Darboux-Halphen system, the same conserved quantity holds:

\vspace{-0.25cm}
\begin{equation}
\label{consvd} Q = \sum_i \bigg( \frac{\omega_i^2}{x_j x_k} \bigg) = \frac{\omega_1^2}{x_2 x_3} + \frac{\omega_2^2}{x_3 x_1} + \frac{\omega_3^2}{x_1 x_2} = constant
\end{equation}

Thus, the conserved quantity holds even for the general case. Now we shall consider a variable change.

\subsection{Change of variables and conserved quantities}

In \cite{Tod}, Tod showed that the classical Darboux-Halphen equations and the conserved quantity (\ref{consvd}) can be redefined via a variable transformation, where the redefined equations can be reduced to the Painleve-VI equation. Here we shall show what the generalized equations transform into under a similar variable transformation. \\

We can perform a variable change shown as follows:

\vspace{-0.5cm}
\begin{equation}
\begin{split}
\omega_1 &= \Omega_1 \frac{\dot{y}_1}{\sqrt{y_1} \big( 1 - y_1 \big)} = 2 \Omega_1 \sqrt{- x_2 x_3} \\
\omega_2 &= \Omega_2 \frac{\dot{y}_1}{\sqrt{y_1 \big( 1 - y_1 \big)}} = 2 \Omega_2 \sqrt{x_3 x_1} \\
\omega_3 &= \Omega_3 \frac{\dot{y}_1}{y_1 \sqrt{ 1 - y_1 }} = 2 \Omega_3 \sqrt{- x_1 x_2}
\end{split}
\end{equation}

This effectively means that we can have the conserved quantity (\ref{consvd}) re-written as

\vspace{-0.5cm}
\begin{equation}
\label{newcons1} \therefore \hspace{1cm} Q = - 4 \big( \Omega_1^2 - \Omega_2^2 + \Omega_3^2 \big)
\end{equation}

If we set $Q = - 8 \gamma$, we should have

\vspace{-0.4cm}
\begin{equation}
\label{newcons2} 2 \gamma = \Omega_3^2 + \Omega_1^2 - \Omega_2^2
\end{equation}

For this co-ordinate transformation, we have the following dynamical equations

\vspace{-0.4cm}
\begin{align}
\label{neweq1} \big( \Omega_1 \big)' &= \frac{\Omega_2 \Omega_3}{y_1} + \frac{\lambda^2}{4 x_1 y_1 \sqrt{- x_2 x_3}} \\ 
\label{neweq2} \big( \Omega_2 \big)' &= \frac{\Omega_1 \Omega_3}{y_1 (1 - y_1)} + \frac{\lambda^2}{4 x_1 y_1 \sqrt{x_1 x_3}} \\ 
\label{neweq3} \big( \Omega_3 \big)' &= \frac{\Omega_1 \Omega_2}{1 - y_1} + \frac{\lambda^2}{4 x_1 y_1 \sqrt{- x_1 x_2}}
\end{align}

It is evident from (\ref{newcons1}) and (\ref{newcons2}), the redefined conserved quantity is the same as that defined by Tod. Thus, aside from the different dynamical equations (\ref{neweq1}) - (\ref{neweq3}) under a co-ordinate transformation, almost all of Tod's work in \cite{Tod} will apply to the generalized Darboux-Halphen system with common additive terms as well.

\section{Generalized DH system with individual additive terms}

So far we have dealt with the generalized Darboux-Halphen system involving a common additive term. Now we shall see how individual additive terms arise. We shall start by reviewing the reduction process on the self-dual Yang-Mills equation, and see which is the starting configuration that produces the desired result.

\vspace{-0.4cm}
\begin{equation} \label{sdym} 
\begin{split}
F_{ab} &= - \frac12 {\varepsilon_{ab}}^{cd} F_{cd} \\
F_{ab} &= \partial_a A_b - \partial_b A_a - \big[ A_a, A_b \big]
\end{split}
\end{equation}

If $A_0 = 0$ and all $A_i = A_i (t)$ only, then (\ref{sdym}) becomes the Nahm equation \cite{Nahmmono, Nahm} described below

\vspace{-0.4cm}
\begin{equation} \label{nahm} 
\begin{split}
F_{0i} &= \dot{A}_i \hspace{2cm} F_{ij} = - \big[ A_i, A_j \big] \\
\Rightarrow &\hspace{1cm} \dot{A}_i = \frac12 {\varepsilon_i}^{jk} \big[ A_j, A_k \big]
\end{split}
\end{equation}

Now, the $A_i$s are functions from $\mathbb{R}^4$ to a Lie algebra $\mathfrak{g}$

\vspace{-0.4cm}
\begin{equation}
\label{fld} A_i = - M_{ij} (t) O_{jk} X_k
\end{equation}

where $O_{ij}$ is an SO(3) matrix, and $X_i$ are the generators of $\mathfrak{sdiff}(S^3)$ satisfying the relation $\big[ X_i, X_j \big] = \varepsilon_{ijk} X_k$. The matrix $M_{ij}$ is given as a sum of  symmetric components $M_s$ and anti-symmetric components $M_a$

\vspace{-0.4cm}
\begin{align}
\label{mat} M &= M_s + M_a = P M_0 P^{-1} = P (d + a) P^{-1} \\ 
\label{sym1} & d = \left({\begin{array}{ccc}
\omega_{10} & 0 & 0\\
0 & \omega_{20} & 0\\
0 & 0 & \omega_{30}
\end{array} } \right) \\
\label{assym} & a = {\varepsilon_{ij}}^k \tau_{k0} = \left({\begin{array}{ccc}
0 & \tau_{30} & - \tau_{20} \\
- \tau_{30} & 0 & \tau_{10} \\
\tau_{20} & - \tau_{10} & 0
\end{array} } \right)
\end{align}

Note that the symmetric matrix $d$ in (\ref{sym1}) has no off-diagonal terms. For $P = I, \omega_i = \omega_{i0}$ and $\tau_i = \tau_{i0}$, leading to a system of equations with common additive terms. The equation we get on applying (\ref{fld}) and (\ref{mat}) to (\ref{nahm}) is  (see Ablowitz et. al. in \cite{ACH1, AC}):

\vspace{-0.5cm}
\begin{equation}
\label{mateq} \dot{M} = \big( \text{Adj} (M) \big)^T + M^T M - \text{Tr} (M) . M
\end{equation}

On setting $P = I$ and taking the diagonal terms, we get the generalized Darboux-Halphen system equations (\ref{gendh}). The off-diagonal terms taken together give us

\vspace{-0.5cm}
\begin{equation}
\label{teq} \dot{\tau}_i = - \tau_i \big( \omega_j + \omega_k \big) \hspace{0.75cm} \tau_i^2 = \alpha_i^2 \big( \omega_j - \omega_i \big) \big( \omega_i - \omega_k \big)
\end{equation}

This concludes the result of the symmetric part $M_s = d$ without off-diagonal terms, giving the generalized Darboux-Halphen system equations with common additive terms represented by $\lambda^2$. We shall now consider the setting that produces generalized D-H equations with individual additive terms.

\subsection{Symmetric off-diagonal terms}

To obtain individual additive terms to each equation, we will consider $P \neq I$ so that the symmetric matrix $M_s$ has symmetric off-diagonal terms as shown below:

\vspace{-0.4cm}
\begin{equation} \label{sym2}
\begin{split}
M_s &= P . d . P^{-1} = \left({\begin{array}{ccc}
\omega_1 & \sigma_3 & \sigma_2\\
\sigma_3 & \omega_2 & \sigma_1\\
\sigma_2 & \sigma_1 & \omega_3
\end{array} } \right) \\
M_a &= P . a . P^{-1} = \left({\begin{array}{ccc}
0 & \tau_3 & - \tau_2 \\
- \tau_3 & 0 & \tau_1 \\
\tau_2 & - \tau_1 & 0
\end{array} } \right)
\end{split}
\end{equation}

This alters the equations to those with individual terms

\vspace{-0.4cm}
\begin{equation}
\label{gdhi} \dot{\omega}_i = \omega_j \omega_k - \omega_i \big( \omega_j + \omega_k \big) + \lambda_i^2
\end{equation}

If we define new variables as $\theta_i = \sigma_i + \tau_i, \ \phi_i = \sigma_i - \tau_i$, then the matrix $M$ in (\ref{mat}) becomes:

\vspace{-0.4cm}
\begin{equation}
M = 
\left({\begin{array}{ccc}
\omega_1 & \theta_3 & \phi_2 \\
\phi_3 & \omega_2 & \theta_1 \\
\theta_2 & \phi_1 & \omega_3
\end{array} } \right)
\end{equation}

 and the differential equations we have from this matrix are elaborated as:

\vspace{-0.4cm}
\begin{align}
\label{ind1} \dot{\omega}_i &= \omega_j \omega_k - \omega_i ( \omega_j + \omega_k ) - \theta_i \phi_i + \theta_j^2 + \phi_k^2 \\ \nonumber \\
\label{ind2} \dot{\theta}_i &= - \big( \theta_i + \phi_i \big) \omega_i - \big( \theta_i - \phi_i \big) \omega_k + \theta_k \big( \theta_j + \phi_j \big) \\ \nonumber \\
\label{ind3} \dot{\phi}_i &= - \big( \theta_i + \phi_i \big) \omega_i + \big( \theta_i - \phi_i \big) \omega_j + \phi_j \big( \theta_k + \phi_k \big)
\end{align}

Thus, we have effectively formulated the generalized Darboux-Halphen system with individual additive terms. 

If we define $\theta = \theta_3 , \phi = \phi_3$, and set

\vspace{-0.25cm}
\begin{equation} \label{conj} 
\theta_1 = \theta_2 = \phi_1 = \phi_2 = 0 
\end{equation}

then the matrix $M$ from (\ref{mat}) and the equations generated upon applying (\ref{conj}) are given as:

\vspace{-0.25cm}
\begin{equation} 
\label{ca} M = \left({\begin{array}{ccc}
\omega_1 & \theta & 0 \\
\phi & \omega_2 & 0 \\
0 & 0 & \omega_3
\end{array} } \right)
\end{equation}

\begin{equation} \label{aceq}
\begin{split}
\dot{\omega}_1 &= \omega_2 \omega_3 - \omega_1 ( \omega_2 + \omega_3 ) - \phi^2 \\
\dot{\omega}_2 &= \omega_3 \omega_1 - \omega_2 ( \omega_3 + \omega_1 ) + \theta^2 \\
\dot{\omega}_3 &= \omega_1 \omega_2 - \omega_3 ( \omega_1 + \omega_2 ) + \theta \phi \\
\dot{\theta} &= - \big( \theta + \phi \big) \omega_3 - \big( \theta - \phi \big) \omega_2 \\
\dot{\phi} &= - \big( \theta + \phi \big) \omega_3 + \big( \theta - \phi \big) \omega_1 
\end{split}
\end{equation}

These are the equations identical to the example found by Ablowitz et. al. in \cite{ACH}. Now we shall compute the form of the conserved quantity $Q$ for this case.

\subsection{Related conserved quantity}

The symmetric matrix (\ref{sym2}) is related to the symmetric matrix of (\ref{sym1}) by a similarity transformation. This means that we could describe the symmetric off-diagonal terms of $M_s$ in terms of the diagonal terms of $d$, and by reversing the transformation, express the conserved quantity for the Darboux-Halphen system with individual additive terms as in equations (\ref{ind1}) - (\ref{ind3}). Now, we will consider the example (\ref{aceq}) from \cite{ACH}. \\

We start with the matrix $M_0$ from (\ref{mat}) for the system with common additive terms and set $\tau_{10} = \tau_{20} = 0$:

\vspace{-0.25cm}
\begin{equation} \label{mat0}
M_0 = \left({\begin{array}{ccc}
\omega_{10} & \tau_{30} & 0 \\
- \tau_{30} & \omega_{20} & 0 \\
0 & 0 & \omega_{30}
\end{array} } \right)
\end{equation}

We will formulate the matrices $M_s$ in (\ref{sym2}) and $M$ in (\ref{ca}) using the similarity transformation matrix:

\vspace{-0.25cm}
\begin{equation}
P = \left({\begin{array}{ccc}
\cos \gamma & \sin \gamma & 0 \\
-\sin \gamma & \cos \gamma & 0 \\
0 & 0 & 1
\end{array} } \right)
\end{equation}

This will transform matrix (\ref{mat0}) into (\ref{ca}) given by substituting $\alpha = \sin \gamma, \ \beta = \cos \gamma, \ \delta = \omega_{10} - \omega_{20}$ as:

\vspace{-0.25cm}
\begin{equation}
\label{newmat} M = \left({\begin{array}{ccc}
\omega_{10} \alpha^2 + \omega_{20} \beta^2 & - \delta \alpha \beta + \tau_{30} & 0 \\
- \delta \alpha \beta - \tau_{30} & \omega_{10} \beta^2 + \omega_{20} \alpha^2 & 0 \\
0 & 0 & \omega_{30}
\end{array} } \right)
\end{equation}

 Thus, comparing (\ref{newmat}) with (\ref{ca}), the transformation rule between the variables for the original system and the similarity transformed one can be worked out to be:

\vspace{-0.4cm}
\begin{equation} \label{transform}
\begin{split}
\omega_{10} + \omega_{20} &= \omega_1 + \omega_2 = \Sigma \\
\omega_{10} - \omega_{20} &= \pm \sqrt{\big( \omega_1 - \omega_2 \big)^2 + \big( \theta + \phi \big)^2} = \pm \Delta
\end{split}
\end{equation}

\begin{equation} \label{trule}
\omega_{10} = \frac12 \Big( \Sigma \pm \Delta \Big) \hspace{0.5cm} \omega_{20} = \frac12 \Big( \Sigma \mp \Delta \Big) \hspace{0.5cm} \omega_{30} = \omega_3
\end{equation}

Using the transformation rule of (\ref{trule}), we can also compute $x_{10}, x_{20}, x_{30}, x_{i0} = \omega_{j0} - \omega_{k0}$ as shown below:

\vspace{-0.4cm}
\begin{equation} \label{tdiff}
\begin{split}
x_{10} &= \frac12 \big( \Sigma \mp \Delta \big) - \omega_3 \\
x_{20} &= \omega_3 - \frac12 \big( \Sigma \pm \Delta \big) \\
x_{30} &= \pm \Delta
\end{split}
\end{equation}

Thus, we can see that $\pm$ setting swaps between $\omega_{10}$ and $\omega_{20}$, and one can apply (\ref{transform}), (\ref{trule}) and (\ref{tdiff}) to the conserved quantity (\ref{consvd}) to express it in terms of the new variables. Further simplification shows that the conserved quantity is indeed invariant under choice of $\pm$. The conserved quantity $Q$ for (\ref{aceq}) in terms of $\omega_1, \omega_2, \omega_3, \theta, \phi$ is given as:

\vspace{-0.4cm}
\begin{align}
Q  = &\frac{\omega_{10}^2}{x_{20} x_{30}} + \frac{\omega_{20}^2}{x_{30} x_{10}} + \frac{\omega_{30}^2}{x_{10} x_{20}} \nonumber \\
= &\frac{\big( \Sigma + \Delta \big)^2}{2 \Delta \big\{ 2 \omega_3 - \big( \Sigma + \Delta \big) \big\} } - \frac{\big( \Sigma - \Delta \big)^2}{2 \Delta \big\{ 2 \omega_3 - \big( \Sigma - \Delta \big) \big\} } \nonumber \\
& \hspace{1cm} - \frac{\omega_3^2}{\omega_3^2 - \omega_3 \Sigma - \frac14 \big( \theta + \phi \big)^2}
\end{align}

This concludes effectively the case study related to the example (\ref{aceq}) of a generalized conserved quantity that accounts for the system with individual additive terms.

\section{Self-duality for membranes and Integrability}

Nahm equation which form the core of our analysis has far reaching consequences in the membrane paradigm and the underlying integrable structure in surface motion of fluids.
Self-dual membranes were first introduced in \cite{bfs} in (2+1) dim where the on-shell equations of motion are trivial and the membrane theory is a 3-dim topological QFT \cite{Zaikov, Fujikawa}. Here we want to stress the correspondence of self-dual membranes in (4+1)-dim to (3+1)-dim SDYM in large $N$ limit. 

The co-ordinates are more precisely the Yang-Mills potentials of the self-dual membrane, $A_i, \forall i = 1, 2, 3$ are functions of time and internal co-ordinates $( \theta, \phi)$ of the sphere. This system involving commutators between matrices $A_1 (t), A_2 (t), A_3 (t)$ of some Lie Algebras appears in study of monopoles \cite{Nahm, AH}. This system of equations is an integrable system as one can easily determine the corresponding linear system for which (\ref{nahm}) serves as a compatibility condition. \\

We write (\ref{nahm}) in this manner:

\vspace{-0.4cm}
\begin{align}
\label{n+} \dot{A}_+ &= i \big[A_-, A_3 \big] \\
\label{n-} \dot{A}_- &= i \big[ A_3, A_+ \big] \hspace{0.5cm} \text{where} \hspace{0.5cm} A_\pm = A_1 \pm i A_2 \\
\label{n3} \dot{A}_3 &= i \big[ A_+, A_- \big]
\end{align}

Equations (\ref{n+}) - (\ref{n3}) have numerous constants of motion, namely $\int A_\pm (t, \theta, \phi)$ which is constant in time, or $\int A_3 A_\pm$ which is also conserved. Another interesting conserved quantity associated with the \textquotedblleft center of mass \textquotedblright of a self-dual membrane is $\int A_i (t, \theta, \phi) = const$.\\

The linear system (\ref{n+}) - (\ref{n3}) can be re-written as

\vspace{-0.4cm}
\begin{equation}
\begin{split}
\dot{\Psi} &= i \big[ A_3 + \lambda A_-, \Psi \big] \\
\dot{\Psi} &= i \big[ \lambda^{-1} A_+ - A_3, \Psi \big]
\end{split}
\end{equation}

for any arbitrary complex $\lambda \neq 0$ such that

\vspace{-0.4cm}
\begin{equation} \label{linear}
\begin{split}
\dot{\Psi} = \mathcal{L}_{A_3 + \lambda A_-} & \Psi \hspace{1cm}
\dot{\Psi} = \mathcal{L}_{\lambda^{-1} A_+ - A_3} \Psi \\
\text{with } \hspace{0.5cm} \mathcal{L}_f = i & \bigg( \frac{\partial f}{\partial \phi} \frac{\partial \ }{\partial \cos \theta} - \frac{\partial f}{\partial \cos \theta} \frac{\partial \ }{\partial \phi} \bigg)
\end{split}
\end{equation}

The compatibility condition (\ref{linear}) is equivalent to 

\vspace{-0.4cm}
\begin{equation}
\label{eqcomp} \big[ \partial_t - \mathcal{L}_{A_3 + \lambda A_-}, \partial _t - \mathcal{L}_{\lambda^{-1} A_+ - A_3} \big] = 0
\end{equation}

It is straightforward to see if one compares the co-efficients of the power of $\lambda$ in (\ref{eqcomp}) namely $\lambda^{-1}, \lambda^0 \text{and } \lambda$ respectively, one ends up with (\ref{n+}), (\ref{n-}) and (\ref{n3}) in that order. In principle, using inverse scattering method, one could in principle construct all solutions of the self-duality equations (\ref{n+}), (\ref{n-}) and (\ref{n3}) from the linear system (\ref{linear}) \cite{BZ, Yang}.

Lastly, in order to obtain conserved quantities for the $SU(\infty)$ Nahm equations (\ref{nahm}) it is not difficult to check that
one can represent the original set of equations  (\ref{n+}) - (\ref{n3})  in Lax form on the Poisson-Lie algebra of functions as
\begin{equation}
\begin{split}
\mathcal{L} &= \frac{1}{\lambda} (A_+ + i A_-) +  \lambda (A_+ - i A_-) - 2iA_3, \Psi \big] \\
\mathcal{M} &= i A_3 -  \lambda (A_+ - i A_-) \\
\dot{\mathcal{L}} &= \big[\mathcal{L}, \mathcal{M}\big]
\end{split}
\end{equation}

From this one can write down the time dependent conserved densities as homogeneous polynomials expressed in integral form
which further leads to the possibility of representing axially symmetric surface motions of relativistic charged spherically symmetric fluids discussed in \cite{halburd}, which we postpone for a future work.

\section{Conclusion}

So far, we have managed to show that the conserved quantity $Q$ from (\ref{consvd}) obtained for the classical case applies even for the generalized Darboux-Halphen system with common additive terms. Furthermore, we have formulated the fully generalized Darboux-Halphen system in equations (\ref{ind1}) - (\ref{ind3}), of which, a specific case was an example worked out in \cite{AC}. Consistency of the form of the conserved quantity from the classical to general case is easily permitted by the common additive term $\lambda^2$ in all three Darboux-Halphen equations. If instead we were to take a different additive term for each of the equations in (\ref{gendh}) as in (\ref{gdhi}), the whole task of producing corresponding conserved quantity becomes a complicated process as (\ref{xprop1}) cannot be directly applied then. 

However, the equations with individual additive terms can be connected to the equations with common additive terms via a similarity transformation that leaves the matrix differential equation invariant as can be seen from (\ref{mat}). This connection would allow us to deduce the form of the conserved quantity for the case with individual additive terms. This further implies that Tod's result could possibly be extrapolated to more general quadratic differential equations of the form 

\vspace{-0.25cm}
\begin{equation}
\dot{X}_i = \sum_{j, k} {a_i}^{jk} X_j X_k
\end{equation}

 introduced in \cite{yoso} and \cite{RM}.

Only one example has been worked out so far for conserved quantities related to equations with different additive terms, 
but other similarity transformations might  be possible that help us describe it in more general way. 
The method employed to study integrability of membranes can be further extended possibly in a future work to study exactly solvable models of relativistic charged fluids. One can further consider the deformation of the symplectic structure to study dissipative fluid dynamics in this membrane paradigm, as well.

\section*{Acknowledgement}
The research of RR was supported by FAPESP through Instituto de Fisica, Universidade de Sao Paulo with grant number 2013/17765-0. RR is grateful to DFMA, IFUSP for the hospitality and support.


\begin{thebibliography}{999}

\bibitem{Tod} K.P. Tod, {\em Self–dual Einstein metrics from the Painlev\'e VI equation}, Phys. Lett. A 190 (1994) 221-224.

\bibitem{Da} G. Darboux, {\em Lecons sur les syst\'emes othogonaux}, (2nd ed.). Gauthiers-Villars, Paris (1910)

\bibitem{Ha} G.H. Halphen, {\em Sur certains syst\'emes d’\'equations diff\'erentielles}, C. R. Acad. Sci. Paris, 92 (1881) 1404-1406. 

\bibitem{AH} M. Atiyah and N. Hitchin, {\it The Geometry and dynamics of magnetic monopoles}, (Princeton University Press, Princeton, NJ, 1988).

\bibitem{GP} G.W. Gibbons and C.N. Pope, {\em The Positive Action Conjecture and Asymptotically Euclidean Metrics
in Quantum Gravity}, Commun. Math. Phys. 66 (1979) 267-290. 

\bibitem{Hi} N. Hitchin, {\em Twistor Spaces, Einstein metrics and isomondromic deformations},
J. Diff. Geom. 42 (1995) 30-112. 

\bibitem{Du}  B. A. Dubrovin, {\em Geometry of 2D topological field theories}, Lecture Notes in Math. 1620, Springer-Verlag, Berlin, Heidelberg, New York (1996), arXiv: hep-th/9407018.

\bibitem{AC} M. J. Ablowitz, S. Chakravarty and R.G. Halburd, {\em Integrable systems and reductions of the self-dual 
Yang-Mills equations}, J. Math. Phys. 44 (2003) 3147.

\bibitem{cgr} S. Chanda, P. Guha, R. Roychowdhury, { \it Bianchi-IX, Darboux-Halphen and Chazy-Ramanujan}, Int. J. Geom. Methods Mod. Phys. {13} (2016), no. 4, 1650042, 25 pp, arXiv:1512.01662 [hep-th].

\bibitem{fintgradflo} S. Chakravarty and R.G. Halburd, { \it First integrals and gradient flow for a generalized Darboux-Halphen system}, Contemp. Math., vol. 301, Amer. Math. Soc., Providence, RI, 2002, 273-281.

\bibitem{ACH} M.J. Ablowitz, S. Chakravarty and R.G. Halburd, {\em On Painlev\'e and Darboux-Halphen type equation},
Proc. 4th Int. Conf. on Mathematical and Numerical Aspects of Wave Propagation. 1998.

\bibitem{Nahmmono} W. Nahm, {\it Monopoles in Quantum Field Theory}, (Trieste, 1981) Pg. 87.

\bibitem{Nahm} W. Nahm, {\em  All self dual  multimonopole for arbitrary  gauge group} CERN, preprint TH. 3172., NATO ASI, B 82 (1983) pp.301

\bibitem{ACH1} M.J. Ablowitz, S. Chakravarty and R.G. Halburd, {\em The generalized Chazy equation from the self-duality equations}, Stud. Appl. Math. 103 (1999)75-88.

\bibitem{bfs} B. Biran, E.G. Florates and G.K. Savvidy, { \it The self-dual closed bosonic membranes}, Phys. Lett. B 198 (1987) 329.

\bibitem{Zaikov} A.P. Zaikov, { \it Self-duality in the theory of the bosonic p-branes}, Phys. Lett. B 211 (1988) 281.

\bibitem{Fujikawa} K. Fujikawa, { \it Comment on p-dimensionally extended objects in (p+1)-dimensional space-time}, Phys. Lett. B 213 (1988) 425.

\bibitem{BZ} A. Belavin and V.E. Zakharov, { \it Yang-Mills equations as inverse scattering problem}, Phys. Lett. B 73 (1978) 53.

\bibitem{Yang} C.N. Yang, { \it Condition of self-duality for SU(2) gauge fields on Euclidean four-dimensional space}, Phys. Rev. Lett. 38 (1977), 1377.

\bibitem{halburd} R. Halburd, { \it Solvable models of relativistic charged spherically symmetric fluids}, Class. Quantum. Grav. 18 (2001), 11-25.

\bibitem{yoso} Y. Ohyama and S. Okumura, {\em Darboux-Halphen-Brioshi system with rank four}, math/0304069v1 [math.CA].

\bibitem{RM} R. S. Maier, {\em Quadratic Differential Systems and Chazy Equations, I} arXiv:1203.0283 [math.CA].



\end{thebibliography}
\end{document}